\documentclass[published]{JHEP}
\JHEP{07(1998)001}

\usepackage{epsfig}
\usepackage{amsmath}
\usepackage{latexsym}

\newcommand\apj[3]{{\it Astrophys.\ J.\ }{\bf #1} (#2) #3} 
\newcommand\slashchar[1]{#1\!\!\!/}
\title{Three-body annihilation of neutralinos below two-body
thresholds}
\author{Xuelei Chen and Marc Kamionkowski \\
Department of Physics, Columbia University,\\
538 West 120th Street New York, NY~~10027, USA\\ 
Email: \email{xuelei@phys.columbia.edu}, 
\email{kamion@phys.columbia.edu}}

\received{May 20, 1998} \accepted{July 7, 1998}

\abstract{We calculate the cross section for 
$s$-wave neutralino annihilation to 
three-body final states below the $W^{+} W^{-}$ and 
$t\bar{t}$ thresholds. Such three-body
channels may dominate the annihilation cross section if the neutralino
mass is not too much less than $m_{t}$ and $m_{W}$
respectively. Furthermore, because neutrinos produced in these
channels are much more energetic than those from the $b\bar{b}$ or
$\tau^{+} \tau^{-}$
channels, they can dominate the energetic-neutrino fluxes from
neutralino annihilation in the Sun or Earth far below these thresholds 
and significantly enhance the neutrino signal in certain regions of the 
supersymmetric parameter space. 
}

\begin{document} 

\long\def\comment#1{}
\def\VEV#1{{\left\langle #1 \right\rangle}}

\section{Introduction}

It has long been well established that the observed luminous matter
in Galactic halos cannot account for their total mass.  Determination of
the identity of this unseen dark matter has become one of the most
important problems in modern cosmology. Perhaps the most promising 
dark-matter candidate is the neutralino $\chi$ \cite{report}, a
linear combination of the supersymmetric partners of the $Z$, $\gamma$,
and Higgs bosons.

The existence of neutralinos in our halo could be inferred by
observation of energetic neutrinos from annihilation of
neutralinos that have accumulated at the core of the Sun and/or
Earth in detectors such as AMANDA,
super-Kamiokande, MACRO, and HANUL \cite{energeticneutrinos}.  
These neutrinos are produced
by decays of neutralino annihilation products such as $\tau$ leptons, 
$c$, $b$, and $t$ quarks, and gauge and Higgs bosons if the
neutralino is heavy enough.  In all cases where the signal is
expected to be observable by these or subsequent-generation
detectors, accretion of neutralinos from the Galactic halo onto
the Sun or Earth comes into equilibrium with their depletion
through annihilation \cite{kamionmodind}. Therefore, the annihilation rate depends
on the rate of capture of neutralinos from the halo.  

Although the total annihilation cross section is not needed for
flux predictions, the branching ratios into the various
annihilation products are:  The rate for observation of
neutrinos from decays of various annihilation products may
differ considerably. Energetic neutrinos are much more easier to 
detect than low-energy ones. For example, the flux of energetic
neutrinos from decays of $b$ quarks is much smaller than that
from gauge bosons or top quarks with the same injection energy.

The best technique for inferring the existence of these neutrinos is 
to observe an upward muon produced by a charged-current
interaction in the rock below the detector. The rate for observation
of energetic neutrinos is proportional to the second moment of the 
neutrino energy spectrum, so it is this neutrino energy
moment weighted by the corresponding branching ratio that determines the
detection rate.

By now, the cross sections for annihilation have been calculated
for all two-body final states that arise at tree level
(fermion-antifermion and gauge- and Higgs-boson pairs). 
The precise branching ratios depend on numerous
coupling constants and superpartner masses.  However, roughly
speaking, among the two-body channels the $b\bar{b}$ and
$\tau^{+}\tau^{-}$ 
final states usually dominate for $m_\chi < m_W$.  
Neutralinos that are mostly higgsino annihilate
primarily to gauge bosons if $m_\chi>m_W$,  because there is no
$s$-wave suppression mechanism for this channel.
Neutralinos that are mostly gaugino continue to annihilate primarily
to  $b\bar{b}$ pairs until the neutralino mass exceeds the top-quark mass,
after which the $t\bar t$ final state dominates, as the cross section for
annihilation to fermions is proportional to the square of the fermion mass.  

Three-body final states arise only at higher order in
perturbation theory and are therefore usually negligible.
However, as we already noted, some two-body channels 
easily dominates the cross section when they are open because of 
their large couplings, for example the $W^{+} W^{-}$ for the higgsinos and
$t \bar{t}$ for gauginos.
This suggests that their corresponding three-body 
final states can be important just below these thresholds. 

More importantly, as mentioned above, the rate for indirect detection is also
proportional to the second moment of the energy of the muon neutrino.
The neutrinos produced in these three-body final states are generally
much more energetic than those produced in $b$ and $\tau$ decays, 
so even when the cross sections to these channels are small compared 
with others, they could dominate the indirect-detection rate.

In this paper, we calculate the cross section for the processes
$\chi \chi \to W^{+} W^{-*} \to W f \bar{f'}$ and 
$\chi \chi \rightarrow t\bar t^* \rightarrow t W^{-} \bar{b}$, 
and their charge conjugates (heretofore referred as $tWb$ and $WW^*$ states)
in the $v_{\rm rel} \to 0$ limit, where the star denotes virtual particles. 
Our calculation for the $W W^* $ is applicable for generic
neutralinos, but it is mostly significant for neutralinos that are
primarily made of higgsino. This is because the gauginos have small 
couplings to $W^{+} W^{-}$ pairs. 
As for the $t W b$ final state, our calculation is only applicable to
the gaugino.  If the neutralino is primarily
higgsino, then annihilation to this final state could
also proceed through a $WW^*$ intermediate state, which we have not
included in our calculation.  However, if the neutralino is primarily
higgsino with $m_{W}<m_{\chi}<m_{t}$, then it will annihilate
primarily to $W^{+} W^{-}$ pairs below the $t \bar{t}$
threshold. Furthermore, the neutrino yield from $W^{+} W^{-}$ pairs is
similar to that for $t \bar{t}$ pairs. Therefore, neutrino rates are
affected only weakly if the process $\chi\chi\to W W^{*} \to t W b$
is neglected.

Annihilation in the Sun and Earth takes place
only when the annihilating neutralinos have relative
velocities $v\sim 10^{-3} \ll 1$.  Therefore, if we expand the
annihilation cross section as
\begin{equation}
     \sigma v = a + b v^2\,,
\label{vpower}
\end{equation}
then we consider only the $a$ (the $s$-wave) term.
The relic abundance also depends on the annihilation cross
section.  However, annihilation in the early Universe takes
place when $v\sim0.5$.  In the early Universe, thermal averaging
of the cross section smoothes out the jump in the annihilation
rate near particle thresholds \cite{kimpaolo}.  Therefore, three-body
final states should have only a negligible effect on the relic
abundance, so we do not consider it further.

We calculate the cross section and neutrino signal for the $t t^*$ final
state in Section II and those for
the $W W^*$ final state in Section III, and discuss these results in
Section IV. Some lengthy matrix elements needed for
the calculation are given in the Appendices.

\section{Calculation of the $t W b$ final state}

\EPSFIGURE[b]{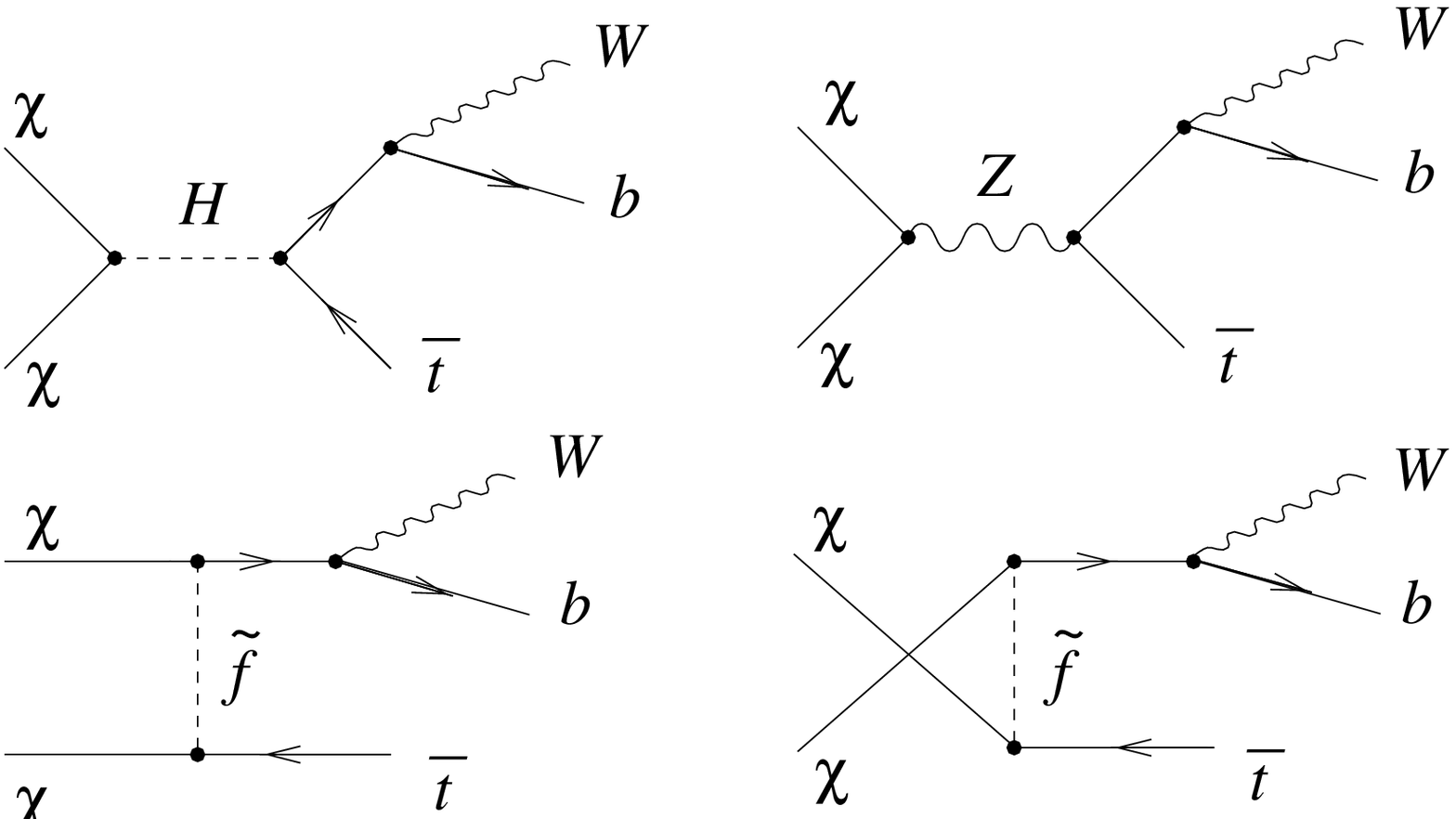, width=5in}
{Feynman diagrams for annihilation of neutralinos to the
$tW^{-}\bar b$ final state through an intermediate $t\bar t^*$ 
state.\label{Feynman_twb}}

We begin by calculating the cross section to the $tWb$
final state.  In some respects, our calculation parallels 
a recent calculation of the three-body decay of Higgs bosons
into this final state \cite{3body}.  The Feynman diagrams for
this process are shown in Fig.~\ref{Feynman_twb}.  Like annihilation
to $t\bar t$ pairs, annihilation to this three-body final state
takes place via $s$-channel exchange of $Z^0$ and $A^0$
(pseudo-scalar Higgs) and $t$- and $u$-channel exchange
of squarks.  Although there are additional diagrams for this
process, such as those shown in Fig.~\ref{Feynman_twb_omit}, these are
negligible for the regions of parameter space that we
investigate for the reasons discussed in the Introduction.

\EPSFIGURE{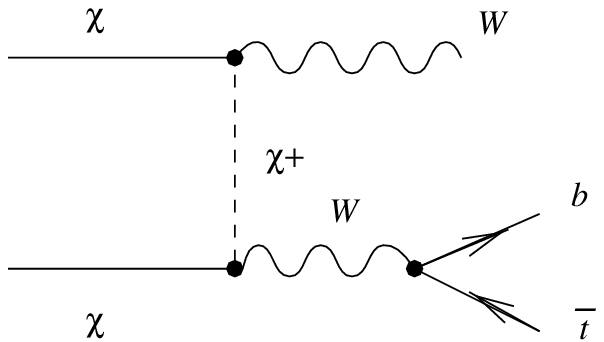,width=4in}
{The contribution to the $tWb$ final state from $W^+W^-$
pairs. These are negligible for the regions of parameter
space we investigate.\label{Feynman_twb_omit}}

The total cross section for annihilation into three-body final
states is given by \cite{Barger} 
\begin{equation}
\sigma=\int \frac{N_{c}}{2 (2\pi)^{5} 
\lambda^{1/2}(s, m_{\chi}^{2}, m_{\chi}^{2})} 
\frac{1}{4}|{\mathcal M}|^{2} d_{3}PS\,;
\end{equation}
where 
\begin{equation}
     \lambda(x, y, z)=x^2 + y^2 + z^2 - 2 x y - 2 y z - 2 z x\,;
\end{equation}
${\cal M}$ is the matrix element for the process; $N_{c}$ is the color
factor; and $d_{3}PS$
is the three-body phase space which can be written as
\begin{eqnarray}
     d_{3}PS &=& \int
     \delta^{4}(p_{1}+p_{2}-p_{4}-p_{5}-p_{6})
     \frac{d^{3}p_{4}}{2p_{4}^{0}}\frac{d^{3}p_{5}}{2p_{5}^{0}}
     \frac{d^{3}p_{6}}{2p_{6}^{0}} \nonumber \\  
     &=&\frac{\pi^{2} s}{4} dx_{4} dx_{6}\,.
\end{eqnarray}
Here 
\begin{equation}
x_{4} = \frac{2 p_{4}^{0}}{\sqrt{s}}\,;\;\;\;\;\;\;\;\; 
x_{5} = \frac{2 p_{5}^{0}}{\sqrt{s}}\,;\;\;\;\;\;\;\;\; 
x_{6} = \frac{2 p_{6}^{0}}{\sqrt{s}}\,;
\end{equation}
$p_{4}$, $p_{5}$, and $p_{6}$ are the four-momenta of the $t$ quark,
$W$ boson and $\bar b$ quarks respectively, as shown in the Feynman
diagrams; and $\sqrt{s}$ is the total center-of-mass energy.  
We approximate $m_{b}=0$,
and take the $v_{\rm rel} \to 0$ limit where $v_{\rm rel}$ is the
relative velocity of the annihilating neutralinos. The
boundaries of the phase space are given by
\begin{eqnarray}
x_{6\rm min}&=&0;\\
x_{6\rm max}&=&1-\frac{(m_{t}+m_{W})^{2}}{4 m_{\chi}^{2}};\\
x_{4\rm min}&=&\frac{1}{2(1-x_{6})}\left[(2-x_{6})\left(1+\frac{m_{t}^{2}}{4 m_{\chi}^{2}}
-\frac{m_{W}^{2}}{4 m_{\chi}^{2}}-x_{6}\right)\right. \nonumber \\
&  &\left.-x_{6} \lambda^{1/2}\left(1-x_{6},
\frac{m_{t}^{2}}{4m_{\chi}^{2}},
\frac{m_{W}^{2}}{4m_{\chi}^{2}}\right)\right];\\
x_{4\rm max}&=&\frac{1}{2(1-x_{6})}\left[(2-x_{6})\left(1+\frac{m_{t}^{2}}{4 m_{\chi}^{2}}-
\frac{m_{W}^{2}}{4 m_{\chi}^{2}}-x_{6}\right)\right. \nonumber \\
& &\left.+x_{6}\lambda^{1/2}\left(1-x_{6},
\frac{m_{t}^{2}}{4m_{\chi}^{2}},
\frac{m_{W}^{2}}{4m_{\chi}^{2}}\right)\right].
\end{eqnarray}

The total cross section is then given by 
\begin{equation}
\label{phase space}
     \sigma v_{\rm rel} = \frac{N_{c}}{128\pi^{3}}
     \int_{x_{6min}}^{x_{6max}}dx_{6}
     \int_{x_{4min}}^{x_{4max}}dx_{4}\frac{1}{4}|{\cal M}|^{2}\,,
\end{equation}
where the spin-summed amplitude squared is given by: 
\begin{equation}
     |{\cal M}|^{2} = \sum_{{\rm spin}}|M|^{2}\,,
\end{equation} 
with 
\begin{equation}
M=M_{1t} + M_{2t} + M_{3t} + M_{4t}-M_{1u} - M_{2u} - M_{3u} - M_{4u}+
M_{5} + M_{7}\,.
\end{equation}
If we make the approximation $t, u \ll m_{\tilde{f}}^{2}$ (we cannot
make this approximation for $s$ in this case), then the $M_{i}$s are
given by 
\begin{eqnarray*}
M_{1t}&=&\frac{g}{\sqrt{2}} c_{1} \bar{u_{6}} \gamma_{\mu} 
L \epsilon_{5}^{\mu} \frac{1}{\slashchar{p}_{3}-m_{t}}
L u_{1}\bar{v_{2}}R v_{4}\,;  \qquad
M_{1u}=\frac{g}{\sqrt{2}} c_{1} \bar{u_{6}} \gamma_{\mu} 
L \epsilon_{5}^{\mu} \frac{1}{\slashchar{p}_{3}-m_{t}}
L u_{2}\bar{v_{1}}R v_{4}\,;  \\
M_{2t}&=&\frac{g}{\sqrt{2}} c_{2} \bar{u_{6}} \gamma_{\mu} 
L \epsilon_{5}^{\mu} \frac{1}{\slashchar{p}_{3}-m_{t}}
R u_{1}\bar{v_{2}}L v_{4}\,;  \qquad
M_{2u}=\frac{g}{\sqrt{2}} c_{2} \bar{u_{6}} \gamma_{\mu} 
L \epsilon_{5}^{\mu} \frac{1}{\slashchar{p}_{3}-m_{t}}
R u_{2}\bar{v_{1}}L v_{4}\,;  \\
M_{3t}&=&\frac{g}{\sqrt{2}} c_{3} \bar{u_{6}} \gamma_{\mu} 
L \epsilon_{5}^{\mu} \frac{1}{\slashchar{p}_{3}-m_{t}}
L u_{1}\bar{v_{2}}L v_{4}\,;  \qquad
M_{3u}=\frac{g}{\sqrt{2}} c_{3} \bar{u_{6}} \gamma_{\mu} 
L \epsilon_{5}^{\mu} \frac{1}{\slashchar{p}_{3}-m_{t}}
L u_{2}\bar{v_{1}}L v_{4}\,;  \\
M_{4t}&=&\frac{g}{\sqrt{2}} c_{4} \bar{u_{6}} \gamma_{\mu} 
L \epsilon_{5}^{\mu} \frac{1}{\slashchar{p}_{3}-m_{t}}
R u_{1}\bar{v_{2}}R v_{4}\,;  \qquad
M_{4u}=\frac{g}{\sqrt{2}} c_{4} \bar{u_{6}} \gamma_{\mu} 
L \epsilon_{5}^{\mu} \frac{1}{\slashchar{p}_{3}-m_{t}}
R u_{2}\bar{v_{1}}R v_{4}\,;  
\end{eqnarray*}
and
\begin{eqnarray*}
M_{5}&=&\frac{g}{\sqrt{2}} \bar{u_{6}} \gamma_{\mu} 
L \epsilon_{5}^{\mu} \frac{1}{\slashchar{p}_{3}-m_{t}}
(c_{5}R+c_{6}L)\gamma^{\nu}
v_{4}\bar{v_{2}}\gamma_{\nu}\gamma_{5}u_{1}\,;\\
M_{7}&=&\frac{g}{\sqrt{2}} c_{7} \bar{u_{6}} \gamma_{\mu} 
L \epsilon_{5}^{\mu} \frac{1}{\slashchar{p}_{3}-m_{t}}
\gamma_{5} v_{4}\bar{v_{2}} \gamma_{5} u_{1}\,.
\end{eqnarray*}
Here, the subscripts 4,5,6 denotes the $t(\bar{t})$, $W^{-}$($W^{+}$)
and $\bar b$($b$) respectively, and $p_{3}=p_{5}+p_{6}$ is the
4-momentum of the virtual top quark. The subscript $t$ and $u$ denotes
$t$- and $u$-channel exchange of a sfermion. The left- and right-hand
projectors $L$ and $R$ are
\begin{equation}
     R=\frac{1+\gamma_{5}}{2}, \qquad L=\frac{1-\gamma_{5}}{2}\,;
\end{equation}
and the coefficients $c_{i}$ are given by 
\begin{eqnarray*}
    c_{1}&=&\sum_{j=1} ^{6}
    \frac{1}{m_{\tilde{f_{j}}}^{2}} W_{j}^{'2}\,; \qquad 
    c_{2}=\sum_{j=1} ^{6}
     \frac{1}{m_{\tilde{f_{j}}}^{2}} X_{j}^{'2}\,; \qquad
     c_{3}=c_{4}=\sum_{j=1}^{6}
     \frac{1}{m_{\tilde{f_{j}}}^{2}} W'_{j} 
     X'_{j};\\
     c_{5}&=&\frac{2g^{2}c_{r}
     (N_{30}^{2}-N_{40}^{2})}{s\cos^{2}\theta_{w}-4 m_{W}^{2}}\, ;\;\;
     c_{6}=\frac{2g^{2}c_{l}
     (N_{30}^{2}-N_{40}^{2})}{s\cos^{2}\theta_{w}-4 m_{W}^{2}}\, ;\;\;
     c_{7}=\frac{2g T_{A00} h_{Aff}}{s-2 m_{A}^2}\,;
\end{eqnarray*}
where the sum on $j$ is over the squarks; $j=1,2,3$ for
left-handed squarks and $j=4,5,6$ for right-handed squarks.  
The quantities $m_{\tilde{f_{j}}}$ are the masses of
sfermions $j$.  The couplings $W'_{j}$ and $X'_{j}$ are $\chi f
\tilde{f}$ right-hand and left-hand couplings respectively; they
are defined as $W'_{u,t,j,0}$ and $X'_{u,t,j,0}$ in the notation of 
\cite{report}, and $T_{A00}$ and $h_{Aff}$ are couplings of the
Higgs boson $A$. $N_{30}$ and $N_{40}$ are the composition
coefficients of the neutralino given in
\cite{report}.  Also,
\begin{equation}
     c_{l}=\frac{1}{2} - \frac{2}{3} \sin^{2}\theta_{W}\,;\;\;\;\;\;\;\;\;
     c_{r}=- \frac{2}{3} \sin^{2}\theta_{W}\,.
\end{equation}

For the virtual-top propagator we use the relativistic
Wigner-Weisskopf form
\begin{equation}
     \frac{\slashchar{p}_{3} +  m_{t}}{p_{3}^{2}-m_{t}^{2} + i m_{t}
     \Gamma_{t}}\,.
\end{equation} The decay width of the top quark is not yet
measured, so we use the prediction of the standard model, 
\begin{equation}
     \Gamma_{t}=\frac{G_{f} m_{t}^{3}}{8\pi\sqrt{2}}
     \left(1-\frac{m_{W}^{2}}{m_{t}^{2}}\right)^{2}\left(1+2
     \frac{m_{W}^{2}}{m_{t}^{2}}\right).
\end{equation}

We calculate $|{\cal M}|^{2}$ using standard trace techniques.
Notice also that 
\begin{equation}
M_{it,jt} = M_{iu,ju} \equiv M_{i,j}\,, \;\; 
M_{5, it} = -M_{5, iu} \equiv M_{5, i}\,,\;\;
M_{7, it} = -M_{7, iu} \equiv M_{7, i}\,;
\end{equation}
where $i, j = 1,2,3,4$. Thus, we have
\begin{eqnarray}
|{\cal M}|^{2} =& 2\left[M_{1,1}+M_{2,2}+M_{3,3}+M_{4,4}+ 
2(M_{1,2}+M_{1,3}+M_{1,4}+M_{2,3}+M_{2,4}+M_{3,4})\right]\nonumber\\
&-2\left[M_{1t,1u}+M_{2t,2u}+M_{3t,3u}+M_{4t,4u}\right]+M_{5,5}
+M_{7,7}\nonumber\\
&-4\left[M_{1t,2u}+M_{1t,3u}+M_{1t,4u}+M_{2t,3u}
+M_{2t,4u}+M_{3t,4u}\right]\nonumber\\
&+4\left[M_{5,1}+M_{5,2}+M_{5,3}+M_{5,4}
+M_{7,1}+M_{7,2}+M_{7,3}+M_{7,4}\right].
\end{eqnarray}
These $M_{i,j}$ are given in Appendix~\ref{A}.

\TABLE[t]{
\caption{Model parameters for Fig.~\ref{twbcross}}
\begin{tabular}{|l|l|l|}
\hline
$M_{2}=260-400$ GeV&$\mu= 10^4 $ GeV&$m_{A}=10^3 $ GeV\\
\hline
$\tan\beta= 2$&$m_{\tilde{q}}^{2}=10^6 {\rm GeV}^{2}$
&$m_{\tilde{l}}^{2}=10^6 {\rm GeV}^{2}$\\
\hline
\end{tabular}
\label{ttmodel}
}

The three-body cross section for a series of typical models are shown in 
Fig.~\ref{twbcross}. The parameters of the models are given in 
Table ~\ref{ttmodel}. Here we assume the GUT mass relation $M_{1}=\frac{5}{3}
M_{2}\tan^{2}\theta_{W}$, and the mass of the lightest neutralino
varies as $M_{2}$ varies. Note that our choice of the parameter $\mu
\gg M_{2}$ ensures that the neutralino is primarily gaugino.
We have also chosen diagonal and degenerate mass
matrices for the squarks and sleptons. 
 
As expected, the three-body cross section 
approaches to the two-body value above the top threshold 
(we take $m_{t}=180$ GeV). 
Below the top mass it is non-zero but drops quickly.

\EPSFIGURE[t]{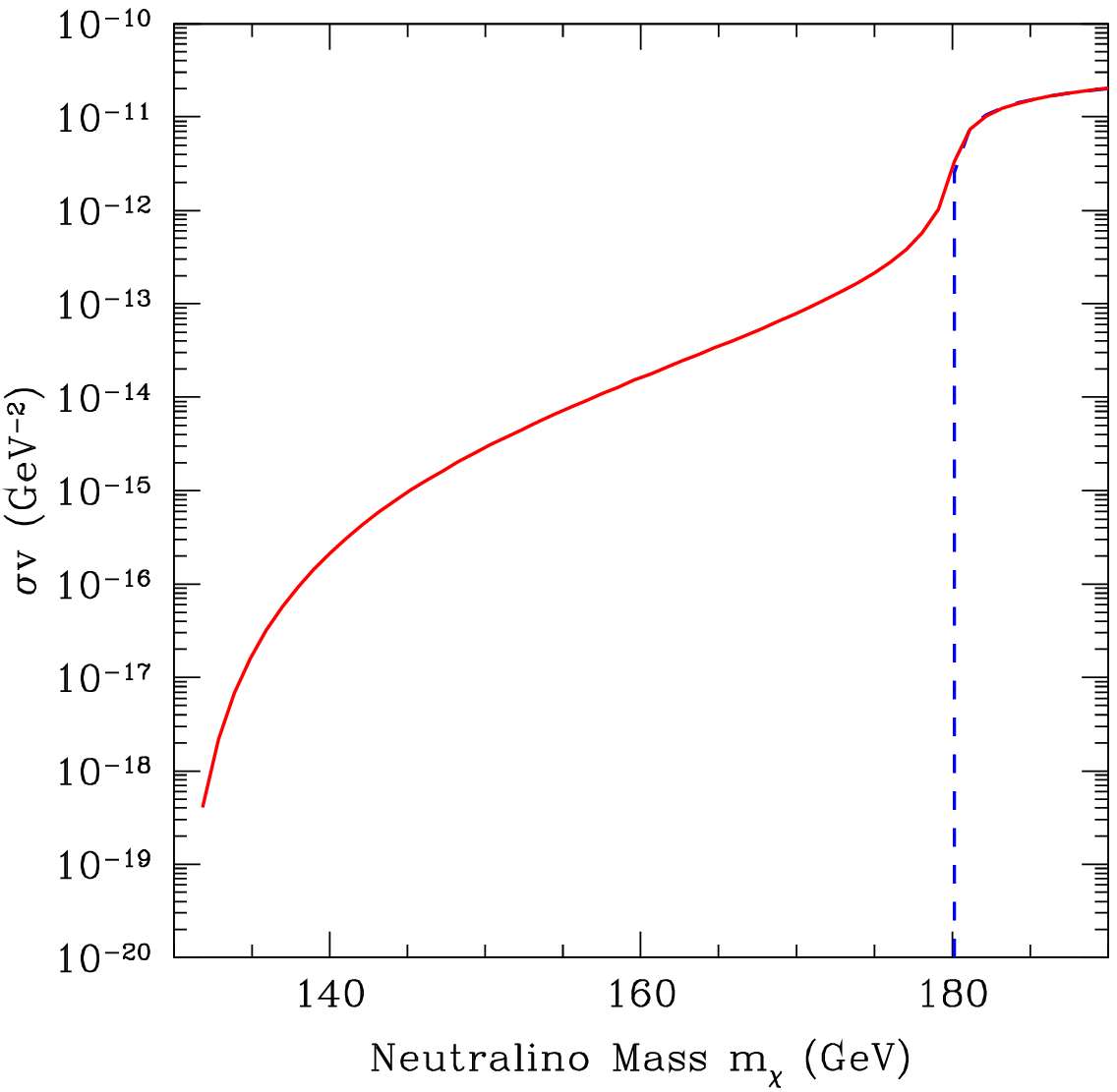}
{The dashed curve is the total annihilation cross section
times relative velocity to two-body final states. The solid curve
includes the $tWb$ final state as well.\label{twbcross}}

\EPSFIGURE[t]{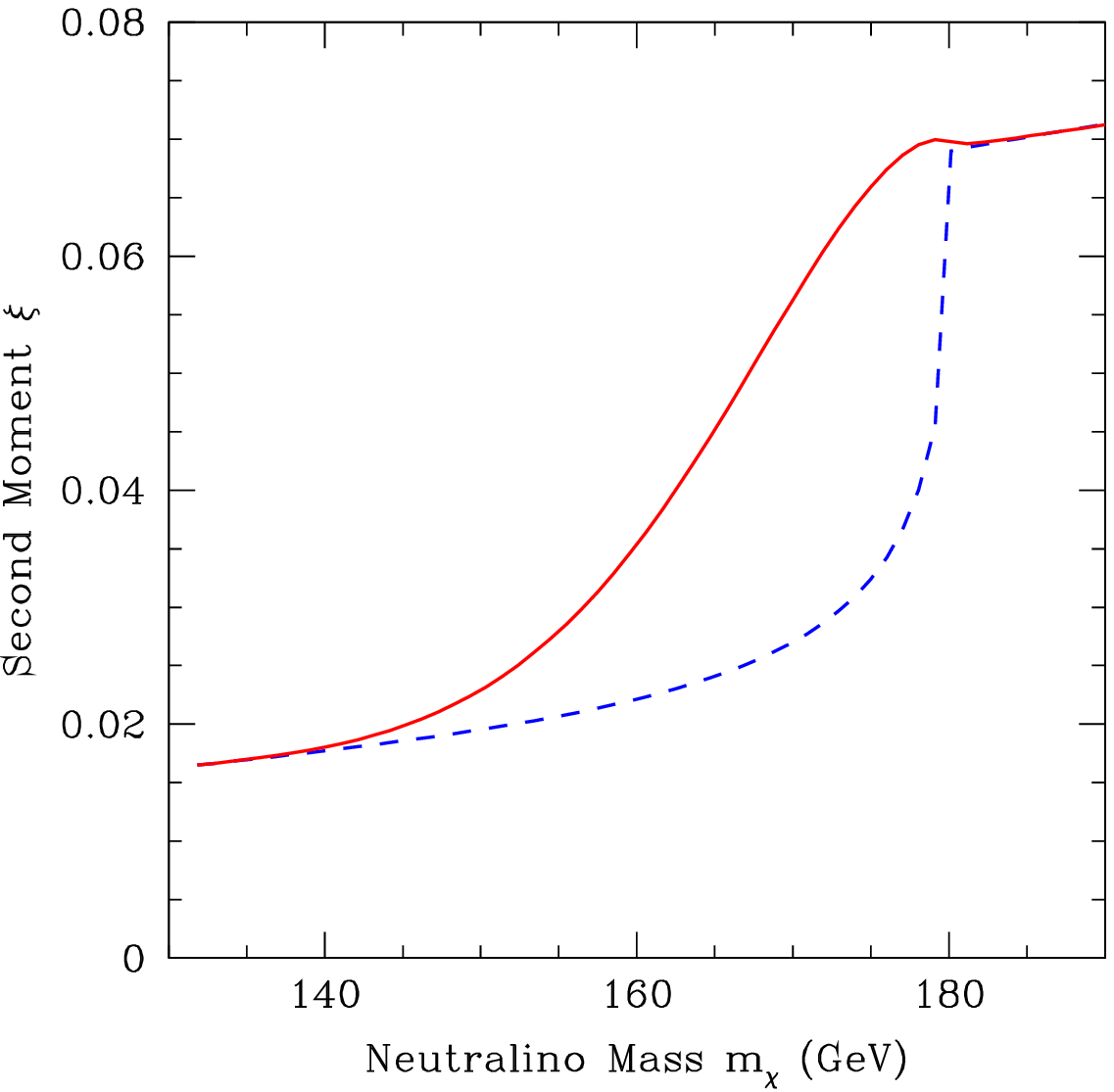}
{Detection rate of neutrinos from neutralino annihilation in
the Earth versus neutralino mass. The ordinate is $\xi$ defined in
Eqs. ~(\ref{eqnxi}) to which the detection rate is proportional.
The dashed curve includes only two-body final states, the solid 
curve includes the $tWb$ final state as well.\label{Nz2figureEarth}}

We now consider the neutrino spectrum. If the rate for neutralino capture in
the Sun is fixed, the flux of upward muon induced by energetic
neutrinos from neutralino annihilation in the Sun and/or Earth depends
on the various neutralino annihilation branching ratios, but not on
the total annihilation cross section.  The flux of upward muons (the
quantity to be compared with experiment) is \cite{report}
\begin{equation}
\Gamma_{\rm detect} \propto \xi_{\rm total} \equiv 
\sum_i a_i b_i \sum_F B_F \VEV{Nz^2}_{F,i}(m_\chi)\,,
\label{Gammarate}
\end{equation}
where the sum on $i$ is over muon neutrinos and anti-neutrinos; $a_i$
are neutrino scattering coefficients, $a_\nu=6.8$ and
$a_{\bar\nu}=3.1$; and $b_i$ are muon-range coefficients, $b_\nu=0.51$
and $b_{\bar\nu}=0.67$ \cite{ritzseckel}.  The sum on $F$ is over
all annihilation channels. The quantity $z=E_\nu/m_{\chi}$ is the
neutrino energy scaled by the neutralino mass. In the two-body case,
this is the same as the injection energy of the particle produced in 
neutralino annihilation, and the second moment is given by
\begin{equation}
\label{Nz2eqn}
\VEV{Nz^2}_{F,i}(E_{\rm in}) \equiv \frac{1}{E_{\rm in}^2} \int
\left( \frac{dN}{dE} \right)_{F,i}(E_\nu,E_{\rm in})~ E_\nu^2 ~dE_\nu\,.
\end{equation}

The functions $\VEV{Nz^2}_{F,i}(E_{\rm in})$ have been calculated 
for injection of all particles that the neutralino may annihilate to
in both the Sun and Earth for both muon neutrinos and 
anti-neutrinos \cite{ritzseckel,neutrino}. 
Note that the $\VEV{Nz^2}$ are different for particles injected in the
Sun and Earth since slowing of quarks and slowing and absorption of
neutrinos may be significant in the Sun but not in the Earth.  For the
same reason, the $\VEV{Nz^2}$ for muon neutrinos and anti-neutrinos
will differ for particles injected into the Sun, although they are the
same for particles injected into the Earth.

In the three-body case, $E_{\rm in} \neq m_{\chi}$. However, 
Eqs.~(\ref{Nz2eqn}) can be generalized. For the $twb$ final state, the
second moment is given by
\begin{equation}
B_F \VEV{Nz^2}=\frac{3}{128\pi^{3}} \int dx_{4} dx_{6}|{\cal M}|^{2} 
\left(\VEV{Nz^2}_{t}x_{4}^{2} 
+\VEV{Nz^2}_{W} x_{5}^{2}
+\VEV{Nz^2}_{b} x_{6}^{2}\right).
\label{eqnxi}
\end{equation}

We plot the result for annihilation in the Earth in 
Fig.~\ref{Nz2figureEarth}. 
(The results for annihilation in the Sun are similar.)
We plot the relevant combination of the second moments $\VEV{Nz^2}$
given in Eq.~(\ref{Gammarate}).  The predicted muon flux is
proportional to this quantity.  The dashed curves in the Figure
show the results obtained by ignoring the $tWb$ final state, and
the solid curves show the results with the $tWb$ final state.

Although the three-body cross section is small except
just below the $t\bar t$ threshold, its contribution to the the 
second moment $B_F \VEV{Nz^2}$ may be important, as illustrated in 
Fig. ~\ref{Nz2figureEarth}. This is because the $\VEV{Nz^2}$ for
top quarks and $W$ bosons is significantly larger than that for the
light fermions \cite{neutrino}.

\section{Neutralino annihilation to $W W^* $}
Below the $W^{+} W^{-}$ threshold, the neutralino can annihilate to a real
$W$ and a virtual $W$ which we denote by $W^*$. The real and virtual 
$W$ bosons then decay independently.  About 10\% of these decay into a
muon (or anti-muon) and a muon anti-neutrino (or neutrino). 
These processes are illustrated in
Fig.~\ref{Feynman_ww}, where the $W^*$ decays into
a fermion pair $f\bar{f'}$, and the fermion pair can be $\tau\nu$,
$\mu\nu$, $e\nu$, $c s$, or $u d$.

The $W W^*$ calculation is similar to the $tWb$ calculation. 
We are only interested in the $v \to 0$
limit. In this limit, only chargino exchange in the $t$ and $u$
channels shown in Fig.~\ref{Feynman_ww} are important.
Neutrinos can be produced either by the
virtual $W^* $ or by decay of the real $W$.

\EPSFIGURE[b]{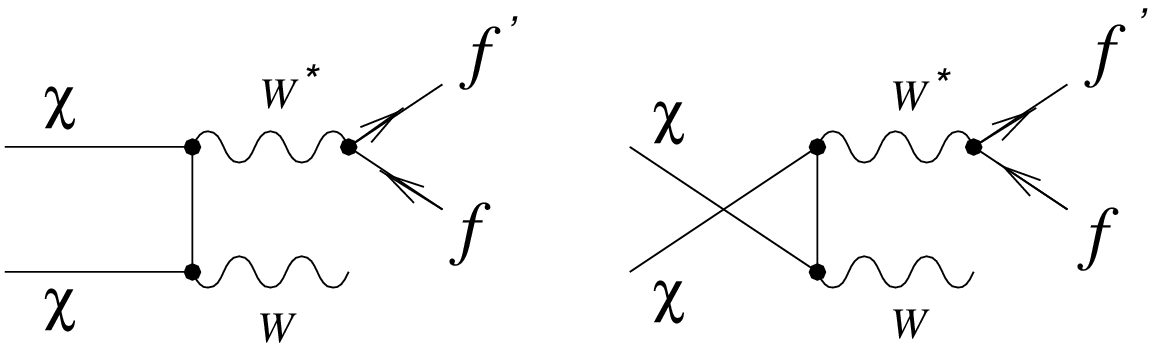,width=6in}
{Neutralino annihilation to $Wf\bar f$ through a virtual $W^*$ by 
chargino exchange in the $t$ or $u$ channel.\label{Feynman_ww}}

The cross section is given again by Eq. (\ref{phase space}), 
and $x_{4}$, $x_{5}$, and $x_{6}$ are defined similarly.  
Neglecting the fermion mass, the boundaries of the phase space 
can be written as:
\begin{eqnarray}
\label{x4eqn1}
x_{4\rm min} &=& \frac{m_{4}}{m_{\chi}},\\
\label{x4eqn2}
x_{4\rm max} &=& 1+\frac{m_{4}^{2}}{4 m_{\chi}^{2}},\\
x_{6\rm min} &=& \frac{1}{2}\left(2-x_{4}-\sqrt{x_{4}^{2}
-\frac{m_{4}^{2}}{m_{\chi}^{2}}}\right),\\
x_{6\rm max} &=& \frac{1}{2}\left(2-x_{4}+\sqrt{x_{4}^{2}
-\frac{m_{4}^{2}}{m_{\chi}^{2}}}\right).
\end{eqnarray}
where $m_{4}=m_{W}$. The amplitude is
\begin{equation}
M=M_{0t}+M_{1t}-M_{0u}-M_{1u}\,,
\end{equation}
where 
\begin{eqnarray}
M_{i,t}&=&\frac{g^2}{k^2 - m_{\chi^{+}_{i}}^2} \bar{v}_{2} \gamma^{\nu} 
(O_{i,L}^2 L +O_{i,R}^2 R)\slashchar{k}\gamma^{\mu} u_{1} W_{4\nu} W_{3\mu}\,,\\
M_{i,u}&=&\frac{g^2}{k^{'2} - m_{\chi^{+}_{i}}^2} \bar{v}_{1} \gamma^{\nu} 
(O_{i,L}^2 L +O_{i,R}^2 R)\slashchar{k}^{'}\gamma^{\mu} u_{2} W_{4\nu} W_{3\mu}\,.
\end{eqnarray}
Here $i=0, 1$ denotes the two chargino states, $m_{\chi^{+}_{i}}$ their
masses, and $O_{i,L}$ and $O_{i,R}$ are the couplings. The quantity
$k=p_{4}-p_{2}$ and $k'=p_{4}-p_{1}$ are the momentum transfered in
the $t$ and $u$ channel respectively, and in the $v \to 0$ limit, $p_{1}=p_{2}$
and $k=k'$. Also, 
\begin{equation}
W_{3\mu}=\frac{g}{\sqrt{2}} \frac{1}{p_{3}^{2} - m_{W}^{2} + i m_{W} 
\Gamma_{W}} \bar{u_{5}} \gamma_{\mu} L u_{1}\,,
\end{equation}
and $W_{4\nu}$ is the wavefunction for the outgoing $W$ boson.

Straightforward calculation yields
\begin{eqnarray*}
|{\cal M}|^{2} = N g^{4}&&\left\{{\displaystyle 
\sum_{i=0}^{1}} \frac{2}{(k^{2}-M_{i}^{2})^{2}}
\left[\left(O_{iL}^{4}+O_{iR}^{4}\right) M_{tta}+ 2 O_{iL}^2
O_{iR}^2 M_{ttb}\right]\right.\\
&&\left.+ \frac{4}{(k^{2}-M_{0}^{2})(k^{2}-M_{1}^{2})}
\left[\left(O_{0L}^{2}
O_{1L}^{2} + O_{0R}^{2} O_{1R}^{2}\right)
\left(M_{tta}-M_{tua}\right)\right.\right.\\
&&\left.\left.+\left(O_{0L}^{2} O_{1R}^{2} + O_{0R}^{2}
O_{1L}^{2}\right)\left(M_{ttb}-M_{tub}\right)\right]\right.\\
&&\left.- {\displaystyle \sum_{i=0}^{1}}\frac{2}{(k^{2}-M_{i}^{2})^{2}} 
\left[\left(O_{iL}^{4}+O_{iR}^{4}\right) M_{tua}+ 2 O_{iL}^2 O_{iR}^2 M_{tub}
\right]\right\}, 
\end{eqnarray*}
where 
\begin{equation}
N=\frac{g^{2} (p_{5}
p_{6})}{(p_{3}^{2}-m_{W}^{2})^{2}+m_{W}^{2}\Gamma_{W}^{2}}\,,
\end{equation}
and the matrix elements $M_{i,j}$ are listed in Appendix~\ref{B}.

This cross section can be evaluated numerically. The result for 
models with parameters given in Table \ref{wwmodel} is shown
in Fig.~\ref{wwcross}. Like the $twb$ case, the three-body
annihilation cross section drops quickly below $m_{W}$
and approaches zero as $m_{\chi} \to m_{W}/2$. 
 
\TABULAR[t]{|l|l|l|}
{\hline
$M_{2}=1000$ GeV&$\mu=55-110$ GeV&$m_{A}=1000$ GeV\\
\hline
$\tan\beta= 2$&$m_{\tilde{q}}^{2}=10^{10} {\rm GeV}^{2}$
&$m_{\tilde{l}}^{2}=10^{10} {\rm GeV}^{2}$\\
\hline}
{Model parameters for Fig.~\ref{wwcross}.\label{wwmodel}}

\EPSFIGURE[t]{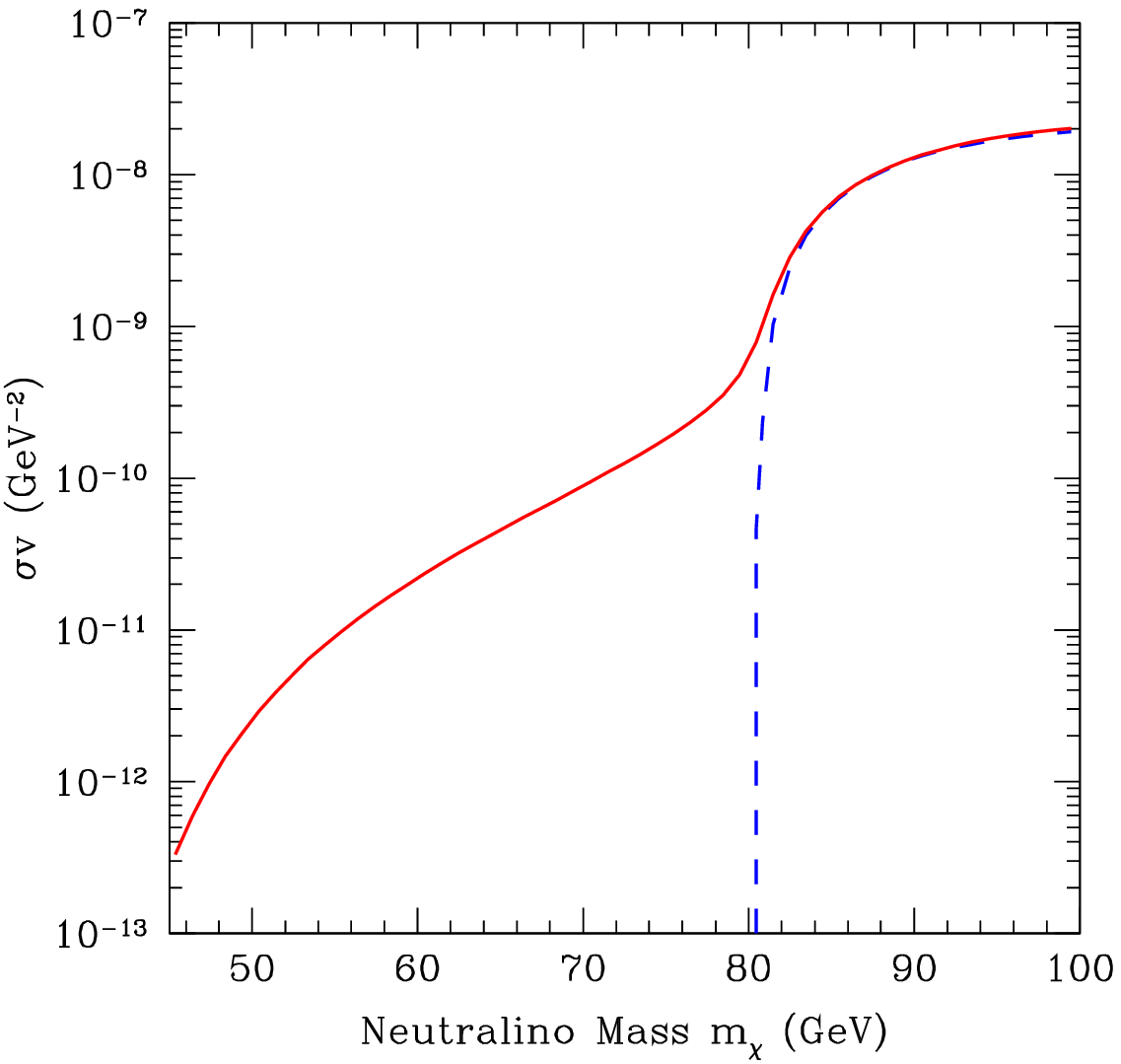}
{Cross section times relative velocity for neutralino annihilation to
$W\mu\nu$. The dashed curve includes only two-body final states and
the solid curve includes the three-body final states.\label{wwcross}}

The calculation of neutrino flux is again analogous to the $twb$ case.
However, we will
not consider the neutrinos produced by decay of the muon or other
fermions as they are of negligibly low energy. We have
\begin{equation}
B_F \VEV{Nz^2}=\frac{1}{128\pi^{3}} \int dx_{4} dx_{6}|{\cal M}|^{2} 
\left(\frac{1}{4}(1+\frac{2}{5}\beta_{W}^{2}) x_{4}^{2} +x_{6}^{2}\right)/2\,,
\label{W3Indirect}
\end{equation}
where $\beta_{W}$ is the velocity of the real $W$ boson. The second term
on the right-hand side accounts for the neutrino produced via the $W^*$. 
The first term is from the decay of the real $W$ boson. This $W$ boson has a
probability of $\Gamma_{W \to \mu\nu}$ to decay to a muon neutrino.
On the other hand this $W$ boson is produced in all $\chi\chi \to W
f\bar{f'}$ channels, and each of these channels has approximately
the same cross section, so the contribution has a weight of 
$n_{chan}\Gamma_{W \to \mu\nu} \approx 1$.

The result for annihilation in the Earth is shown in
Fig.~\ref{W3Earth}. The results for annihilation in the Sun look similar. 
We see again the second moment agrees with the $WW$ result above the
threshold, but does not decrease immediately below $m_{W}$, 
despite the fast drop of the cross section. In fact, 
the value of the second moment is slightly higher below the $WW$ threshold,
and does not drop until the cross section vanishes as $m_{\chi}$ 
approaches $m_{W}/2 $. Below $m_{W}/2$, the four-body channel might 
become significant, and may smooth the jump in much the same fashion
as the three-body channel does near the two-body channel threshold, but
we will not consider it here.

\EPSFIGURE[t]{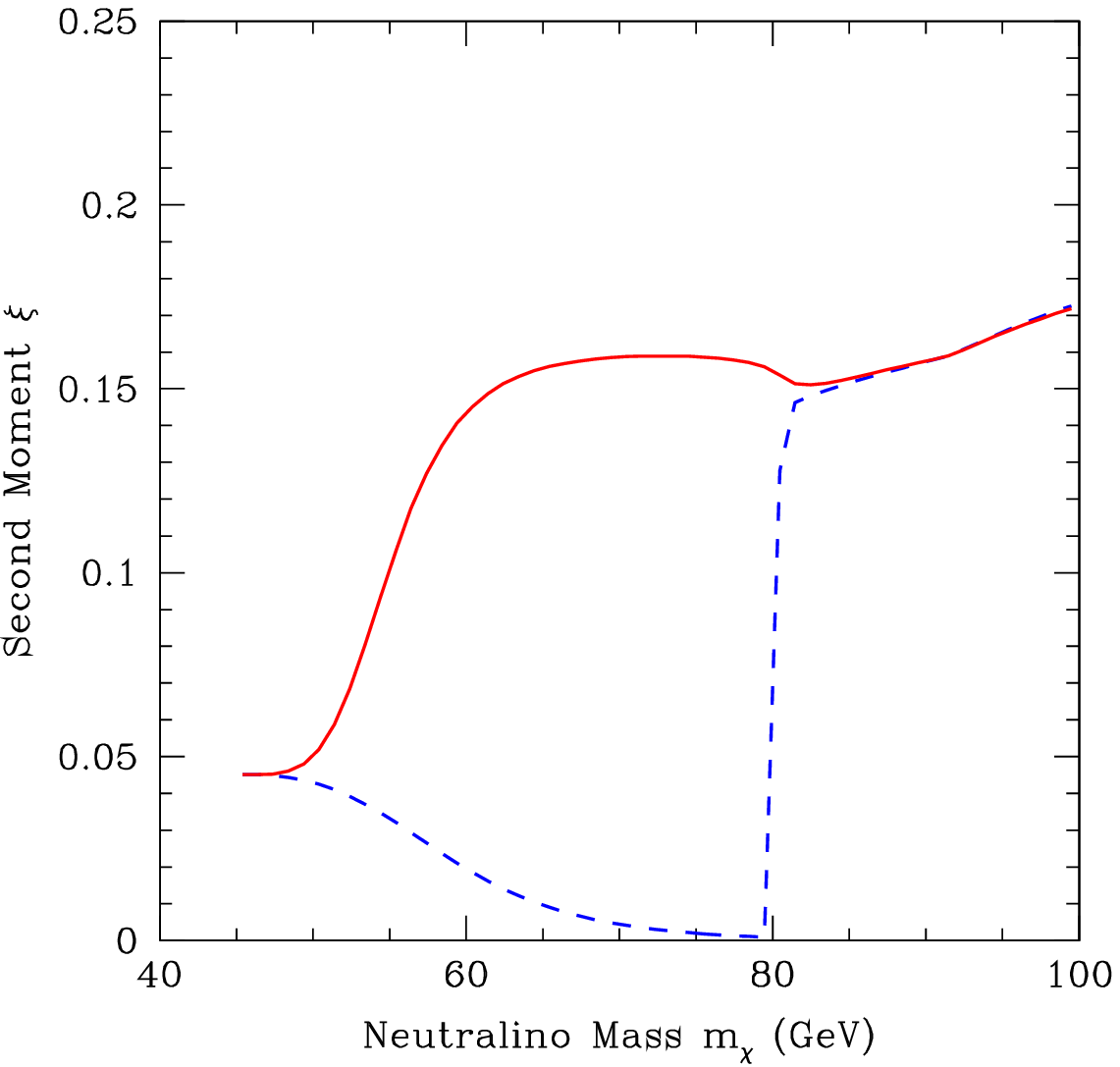}
{Second moment $\xi$ of neutrino energy spectrum for neutrinos
from neutralino annihilation in the Earth. The dashed curve is 
the two-body calculation, and the solid
curve includes the three-body final states from an intermediate $W
W^{*}$ state as well.\label{W3Earth}}

\EPSFIGURE[t]{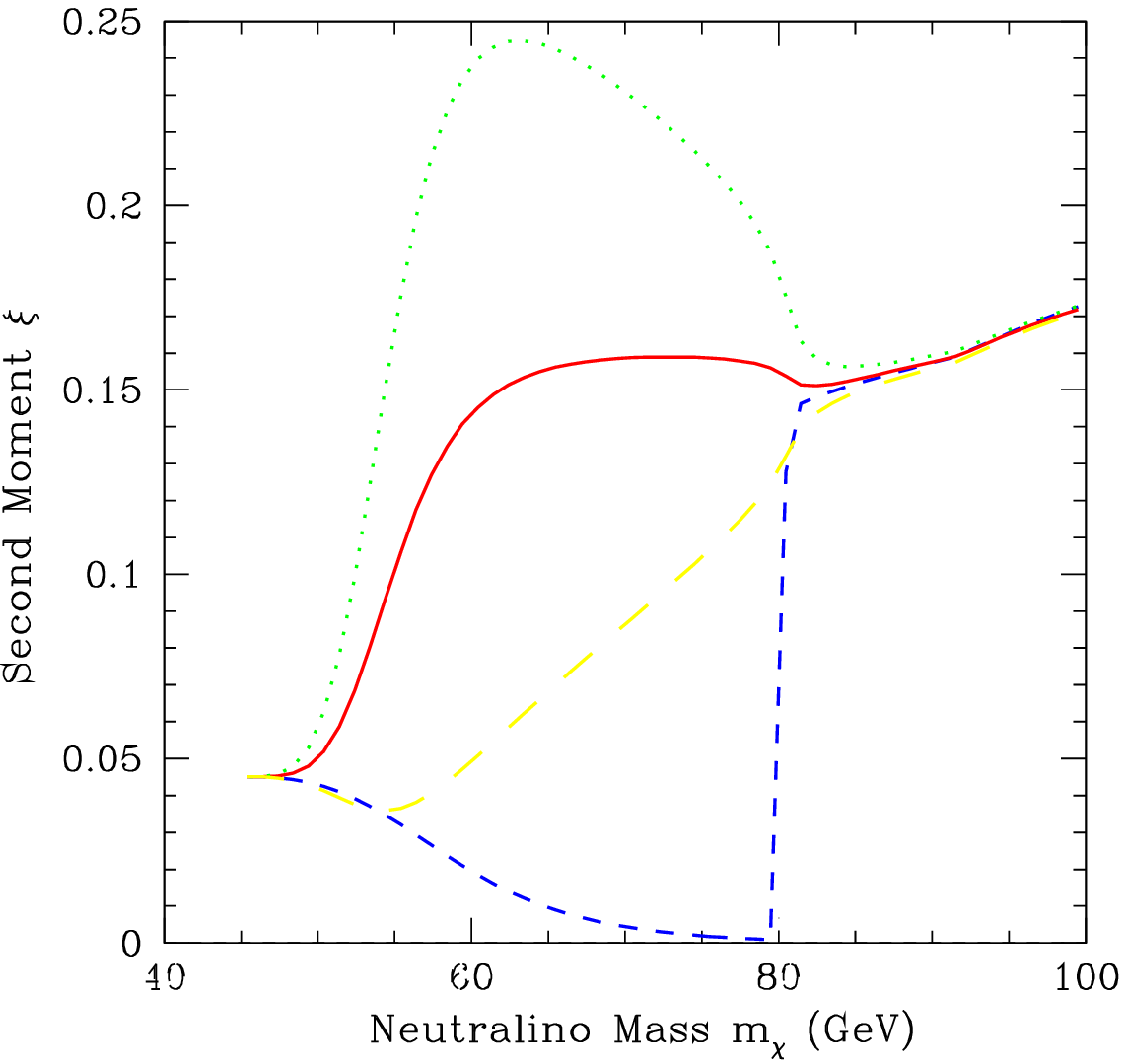}
{Contributions to the second moment $\xi$ for annihilations in
the Earth. Short dashed curve: two-body result; solid curve: three-body
result; dotted curve: (twice of) real $W$ contribution; long dashed
curve: (twice of) virtual $W^{*}$ contribution.\label{W3Anal}\vspace*{-1ex}}

To understand why there is a dip near the $WW$ threshold in 
Fig.~\ref{W3Earth}, consider 
Eq. (\ref{W3Indirect}). There are two terms in the integrand. The
$x_{4}^2$ term is the contribution from the decay of the ``on-shell'' $W$,
and the $x_{6}^{2}$ term is from the virtual $W^{*}$. We plot these
contributions separately in Fig.~\ref{W3Anal}. 

To make comparison easier, what are actually plotted are twice the
contributions from these terms plus contribution from other channels.
As one can see from the plot, both the real and virtual $W$
contributions agree with the two-body result above the threshold as
expected; below the threshold, however, their behaviors are quite
different. While the virtual $W^{*}$ contribution smoothly drops, the
real $W$ contribution actually peaks well below the threshold; then it
decreases as $m_{\chi}$ increases, and has a sharp turn at the $WW$
threshold. This is because the the second moment is proportional to 
$x_{4}^{2}$, and below the threshold $x_{4}$ steadily decreases as
$m_{\chi}$ increases, as can be seen from
Eqs. (\ref{x4eqn1})--(\ref{x4eqn2}). However, above the threshold, $x_{4}$
is stabilized at the $x_{4}=1$ pole, so it stops decreasing, and
gradually increases as $\beta_{W}$ increases. When averaged with the
smooth curve of the $W^*$ contribution, this produces the strange dip at
the threshold.

In this calculation we have assumed that one $W$ is on-shell
while the other is off-shell. It could happen that both $W$'s are
off-shell, and if such contributions are important this may produce a 
smoother curve and the dips in these plots may disappear.  We
examined this possibility by smearing the on-shell mass $m_{4}$ in the range
$m_{W}-3\Gamma_{W} < m_{4} < m_{W}+3\Gamma_{W}$ and calculating an
average with the assumption that the cross section is proportional to 
$$\frac{1}{(m_{4}^{2}-m_{W}^{2})^{2}-m_{W}^{2}\Gamma_{W}^{2}}.$$
This produces a result that differs by less than $10^{-4}$
from the simple on-shell result. Hence it suggests that the on-shell 
approximation works well.

\section{Conclusions}

We made several approximations in our calculation. 
For example, in the $tWb$ calculation, we neglected the 
contributions from diagrams involving production of the $tWb$
final state through virtual $W$ bosons. However, in those regions 
of parameter space where such omissions become unacceptable, the 
process we are considering is not very important.
In the $Wf\bar{f'}$ case, our calculation is generally
applicable, but it is most
useful for a higgsino, since in that case 
the $W^{+} W^{-}$ channel becomes large
above threshold. 

Although the contributions of these three-body channels are 
important only in a limited region of parameter space, it may be a
large effect. In fact, our calculation shows that although the cross
sections of these annihilations are significant only just below the 
two-body channel threshold, due to the high energy of the neutrinos
they produce, they can enhance the neutrino signal by many times and
actually dominate the neutrino signal far below the two-body threshold.  
Furthermore, the regions in question may be of particular interest. For
example, motivated by collider data, Kane and Wells proposed a light
higgsino \cite{light higgsino} (but see \cite{Ellis98}). 
There are also arguments that the
neutralino should be primarily gaugino with a mass somewhere
below but near the top-quark mass \cite{leszek}. 

There are many parameters in the minimal supersymmetric model. The results
shown in Figs.~\ref{Nz2figureEarth} and \ref{W3Earth}
 are of course model dependent, and these effects 
might be more or less important in models with different parameters. 
For a generic neutralino, these effects would be small. 
However, for the regions of parameter spaces we have focussed on 
it seems that the increase
of the indirect-detection rate is fairly robust. For these regions of
parameter space
these effects should be considered for calculations of rates for 
energetic-neutrino detection.

\acknowledgments

We thank G. Jungman for help with the {\tt neutdriver} code.
This work was supported by D.O.E. Contract No. DEFG02-92-ER 40699, 
NASA NAG5-3091, and the Alfred P. Sloan Foundation.

\appendix

\section{The $tt^*$ cross section}\label{A}
We give the $M_{i, j}$ for the $tt^*$ cross section calculation 
in this Appendix.

For convenience, we introduce four-momenta
\begin{equation}
a = -\left[b-\frac{2(p_{3}b)}{p_{3}^{2}} p_{3}\right],\;\;\;\;
b = p_{6} + \frac{2 (p_{5}p_{6})}{m_{W}^{2}} p_{5}\,,
\end{equation}
[these should not be confused with the cross-section factors introduced
in Eq.~(\ref{vpower})]. We further introduce
\begin{equation}
N=\frac{1}{2} \frac{g^{2}}{(p_{3}^{2}-m_{t}^{2})^{2} + \Gamma_{t}^{2}
m_{t}^{2}}\,.
\end{equation}

We can then write the $M_{i, i}$ as
\begin{eqnarray*}
M_{1,1}&=&4 N c_{1}^{2} m_{t}^{2} (p_{1}b)(p_{2}p_{4})\,,\qquad
M_{2,2}=4 N c_{2}^{2} p_{3}^{2} (p_{1}a)(p_{2}p_{4})\,,\\
M_{3,3}&=&4 N c_{3}^{2} m_{t}^{2} (p_{1}b)(p_{2}p_{4})\,,\qquad
M_{4,4}=4 N c_{4}^{2} p_{3}^{2} (p_{1}a)(p_{2}p_{4})\,,
\end{eqnarray*}
and
\begin{eqnarray*}
M_{5,5}=&16 N
&\left\{c_{5}^{2}p_{3}^{2}\left[(p_{1}p_{4})(p_{2}a)+(p_{1}a)(p_{2}p_{4})
-m_{\chi}^{2} (p_{4}a)\right] \right.\\
&&\left. +c_{6}^{2} m_{3}^{2} \left[ (p_{1} p_{4})(p_{2}b)
+ (p_{1}b)(p_{2}p_{4})-m_{\chi}^{2} 
(p_{4}b)\right]\right.\\
&&\left. +2 c_{5} c_{6} m_{t}^{2} (p_{3}b)\left((p_{1}p_{2}) - 
2m_{\chi}^{2})\right)\right\},\\
M_{7,7}=&16 N& c_{7}^{2}
m_{\chi}^{2}\left\{2\left[(p_{3}p_{4})+m_{t}^{2}\right](p_{3}b)
+\left[m_{t}^{2}-p_{3}^{2}\right] (p_{4}b)\right\}.
\end{eqnarray*}
The $M_{i, j}$'s are given by 
\begin{eqnarray*}
     M_{1,2}&=&4 N c_{1} c_{2} m_{\chi}^{2} m_{t}^{2} (p_{3}b)\,,\qquad\qquad\;\;
     M_{1,3}=4 N c_{1} c_{3} m_{\chi} m_{t}^{3} (p_{1}b)\,,\\
     M_{1,4}&=&4 N c_{1} c_{4} m_{\chi} m_{t} (p_{3}b)(p_{2}p_{4})\,,\qquad
     M_{2,3}=4 N c_{2} c_{3} m_{\chi} m_{t} (p_{3}b)(p_{2}p_{4})\,,\\
     M_{2,4}&=&4 N c_{2} c_{4} m_{\chi} m_{t} (p_{3}p_{3})(p_{1}a)\,,\qquad
     M_{3,4}=4 N c_{3} c_{4} m_{\chi}^{2} m_{t}^{2} (p_{3}b)\,,
\end{eqnarray*}
and
\begin{equation*}
\begin{split}
M_{5,1}=-4 N c_{1} m_{t}^{2}&\left\{c_{5}\left[(p_{1}p_{2})(p_{3} b)
+(p_{2}p_{3})(p_{1}b)-(p_{1}p_{3})(p_{2}b)
-2m_{\chi}^{2}(p_{3}b)\right]\right.\\
&\left.+c_{6}\left[2(p_{2}p_{4})(p_{1}b)
-m_{\chi}^{2}(p_{4}b)\right]\right\},
\end{split}
\end{equation*}
\begin{equation*}
\begin{split}
M_{5,2}=&4 N c_{2} \left\{c_{5}\left[4(p_{2}p_{4})(p_{1}p_{3})
(p_{3}b)-2(p_{2}p_{4})(p_{1}b)(p_{3}p_{3})\right.\right.\\
&\left.-2m_{\chi}^{2}(p_{3}p_{4}) (p_{3}b)+m_{\chi}^{2}(p_{3}p_{3})(p_{4}b)\right]\\
&\left.+c_{6}m_{t}^{2}\left[(p_{1}p_{2})(p_{3}b)+(p_{1}p_{3})(p_{2}b)
-(p_{2}p_{3})(p_{1}b)-2m_{\chi}^{2}(p_{3}b)\right]\right\},
\end{split}
\end{equation*}
\begin{equation*}
\begin{split}
M_{5,3}=
4 N c_{3} m_{t} m_{\chi}&\left\{2c_{5}\left[(p_{2}p_{4})(p_{3}b)
+(p_{1}p_{4})(p_{3}b)+(p_{1}p_{3})(p_{4}b)-(p_{3}p_{4})(p_{1}b)\right]\right.\\
&\left.
+c_{6}m_{t}^{2}\left[(p_{3}b)-2(p_{1}b)\right]\right\},\\
\end{split}
\end{equation*}
\begin{equation*}
\begin{split}
M_{5,4}=&
-4 N c_{4} m_{\chi} \left\{c_{5}(p_{3}p_{3})\left[
2(p_{2}p_{4})-(p_{1}p_{4})\right]
+c_{6}m_{t}^{2}[(p_{2}p_{3})-2(p_{1}p_{3})]\right\},
\end{split}
\end{equation*}
\begin{eqnarray*}
M_{7,1}&=&2 N c_{1} c_{7} 
m_{\chi}m_{t}\left(2(p_{3}b)[(p_{3}p_{4})+m_{t}^{2}] 
+(p_{4}b)[m_{t}^{2}-p_{3}^{2}]\right),\\
M_{7,2}&=& 8 N c_{2} c_{7} m_{\chi}^{3} m_{t} (p_{3}b),\qquad
M_{7,3}= 4 N c_{3} c_{7} m_{t}^{2} m_{\chi}^{2} (p_{1}b)\,,\\
M_{7,4}&=& 2 N c_{4} c_{7} \left( p_{3}^2
(p_{1}p_{2})(p_{4}a) + m_{t}^{2}(p_{1}p_{2})(p_{3}b) 
+ m_{t}^{2} m_{\chi}^{2} (p_{3}b)\right),\\
M_{7,5}&=& 8 N c_{7} \left(4 c_{5} m_{t} m_{\chi}^{3} (p_{3}b)
-c_{6} m_{t} m_{\chi} [2((p_{3}p_{4})+m_{t}^{2})(p_{3}b)
+(m_{t}^{2}-p_{3}^{2})(p_{4}b)]\right).
\end{eqnarray*}
\begin{eqnarray*}
M_{1t,1u} &=& 2 N c_{1}^{2} m_{t}^{2} m_{\chi}^{2} (p_{4}b)\,,\qquad\qquad
M_{1t,2u} = 2 N c_{1} c_{2} m_{t}^{2} m_{\chi}^{2} (p_{3}b)\,,\\
M_{1t,3u} &=& 2 N c_{1} c_{3} m_{t}^{3} m_{\chi} (p_{2}b)\,,\qquad\;\quad
M_{2t,2u} = 2 N c_{2}^{2} m_{\chi}^{2} p_{3}^{2} (p_{4}a)\,,\\
M_{2t,4u} &=& 2 N c_{2} c_{4} m_{t} m_{\chi} p_{3}^{2} (p_{2}a)\,,\qquad\;
M_{3t,4u} = 2 N c_{3} c_{4} m_{t}^{2} (p_{1}p_{2}) (p_{3}b)\,,\\
M_{1t,4u} &=& 2 N c_{1} c_{4} m_{t} m_{\chi} [(p_{2}p_{4})(p_{3}b)
	+(p_{2}p_{3})(p_{4}b)-(p_{3}p_{4})(p_{2}b)]\,,\\
M_{2t,3u} &=& 2 N c_{2} c_{3} m_{t} m_{\chi} [(p_{2}p_{4})(p_{3}b)
	+(p_{3}p_{4})(p_{2}b)-(p_{2}p_{3})(p_{4}b)]\,,\\
M_{3t,3u} &=& 2 N c_{3}^{2} m_{t}^{2} [(p_{2}p_{4})(p_{1}b)
	+(p_{1}p_{4})(p_{2}b)-(p_{1}p_{2})(p_{4}b)]\,,\\
M_{4t,4u} &=& 2 N c_{4}^{2} p_{3}^{2} [(p_{2}p_{4})(p_{1}a)
	+(p_{1}p_{4})(p_{2}a)-(p_{1}p_{2})(p_{4}a)]\,.\\
\end{eqnarray*}

In the $v \to 0$ limit, we can approximate
$\vec{p}_{1}=\vec{p}_{2}=0$. 
In terms of $x_{4}$ and $x_{6}$, we have 
\begin{eqnarray*}
(p_{1}p_{2}) &\approx& m_{\chi}^{2}\,,\qquad \qquad\qquad \qquad\qquad
(p_{1}p_{3}) \approx (p_{2}p_{3}) \approx
m_{\chi}^{2} (2-x_{4})\,,\\ 
(p_{1}p_{4})&\approx&(p_{2}p_{4})\approx m_{\chi}^{2}x_{4}\,,
\qquad \qquad \quad
(p_{3}p_{3}) \approx 4 m_{\chi}^{2} (1-x_{4}) + m_{t}^{2}\,,\\
(p_{3}p_{4}) &\approx& 2 m_{\chi}^{2} x_{4} - m_{t}^{2}\,,\qquad\qquad\qquad
(p_{3}p_{5}) \approx 2 m_{\chi}^{2}
(1-x_{4}+\frac{1}{4}\Delta_{456})\,,\\ 
(p_{3}p_{6}) &\approx& 2 m_{\chi}^{2}
(1-x_{4}+\frac{1}{4}\Delta_{645})\,,\qquad
(p_{4}p_{5}) \approx 2 m_{\chi}^{2}
(1-x_{6}-\frac{1}{4}\Delta_{456})\,,\\ 
(p_{4}p_{6}) &\approx& 2
m_{\chi}^{2}(1-x_{5}-\frac{1}{4}\Delta_{645})\,,\qquad 
(p_{5}p_{6}) \approx
2 m_{\chi}^{2}(1-x_{4}-\frac{1}{4}\Delta_{564})\,,
\end{eqnarray*}
where
\begin{equation}
\Delta_{i,j,k}\equiv \frac{m_{i}^{2} + m_{j}^{2} -
m_{k}^{2}}{m_{\chi}^{2}}\,.
\end{equation}

\section{The $WW^*$ cross section}\label{B}
We give the $M_{i, j}$ necessary for the $WW^*$ calculation in this Appendix.

\begin{eqnarray*}
M_{tta}&=&2\left[2 (k q) (p_{2} k) - k^2 (p_{2} q) 
+ \frac{4}{m_{4}^{2}} (p_{2} p_{4}) (p_{4} k) (k q)
- \frac{2}{m_{4}^{2}} (p_{2} p_{4}) (p_{4} q) k^2\right],\\
M_{ttb}&=&-12 m_{\chi}^{2} k^{2}\,,\\
M_{tua}&=&-8m_{\chi}^{2}\left[ k^2 - 
\frac{(p_{4} k)^{2}}{m_{4}^{2}} -\frac{(p_{5} k)
(p_{6} k)}{m_{4}^{2}} - 
\frac{(p_{4} k)(p_{4} p_{5})(p_{6} k)}{m_{4}^{4}}
\right. \nonumber\\
&&-\left. \frac{(p_{4} k)(p_{4} p_{6})(p_{5} k)}{m_{4}^{4}}
+\frac{(p_{4} k)^{2} (p_{5} p_{6})}{m_{4}^{4}}
+\frac{k^2 (p_{4} p_{5})(p_{4} p_{6})}{m_{4}^{4}}
-\frac{k^{2}}{2m_{4}^{2}} (p_{5} p_{6})\right],
\end{eqnarray*}
\begin{eqnarray*}
M_{tub} &=& \frac{1}{2} \left\{ 2\left[-12m_{\chi}^{2} k^{2} + 
8 (p_{1} k)^2
+8\frac{m_\chi^{2}}{m_{4}^{2}}(p_{4} k)^2 
+ 8\frac{(p_{1} p_{4})^2 }{m_{4}^{2}} k^2
-16\frac{(p_{4} k)(p_{1} k)(p_{1} p_{4})}{m_{w}^{2}}\right]\right.\\
&&\left. +\left[k^2 -\frac{2 (p_{5} k)(p_{6} k)}{(p_{5}
p_{6})}\right]\left[8 m_{\chi}^{2}
-16 \frac{(p_{1} p_{4})^2}{m_{4}^{2}}\right]\right.\\
&&\left.+\left[m_{\chi}^2 -\frac{2 (p_{1} p_{5})(p_{1}
p_{6})}{(p_{5} p_{6})}\right]\left[8 k^{2}-16 \frac{(p_{4} k)^2}{m_{4}^{2}}\right]\right.\\
&&\left. +\left[m_{4}^2 -\frac{2 (p_{4} p_{5})(p_{4} p_{6})}{(p_{5}
p_{6})}\right]\left[8\frac{m_{\chi}^{2}}{m_{4}^{2}} k^{2}-16 
\frac{(p_{1} k)^2}{m_{4}^{2}}\right]\right.\\
&&\left.
+2\left[(p_{1} k)-\frac{(p_{1} p_{5})(p_{6} k)
+(p_{1} p_{6})(p_{5} k)}{p_{5} p_{6}}\right]
\left[-8 (p_{1} k)\right]\right.\\
&&\left.
+2\left[(p_{4} k)-\frac{(p_{4} p_{5})(p_{6} k)+
(p_{4} p_{6})(p_{5} k)}{p_{5} p_{6}}\right]\left[16
\frac{(p_{1}p_{4})(p_{1} k)}{m_{4}^{2}}-
8\frac{m_{\chi}^{2}(p_{4} k)}{m_{4}^{2}} \right]\right.\\
&&\left.
+2\left[(p_{1} p_{4})-\frac{(p_{1} p_{5})(p_{4} p_{6})+
(p_{4} p_{5})(p_{1} p_{6})}{p_{5} p_{6}}\right]\left[16
\frac{(p_{1} k)(p_{4} k)}{m_{4}^{2}}-
8\frac{k^{2}(p_{1} p_{4})}{m_{4}^{2}} \right]
\right\}.
\end{eqnarray*}

As in the $tWb$ case, these four-vector products can be
expressed by the variables $x_{4}$, $x_{5}$ and $x_{6}$:
\begin{eqnarray*}
(p_{1} p_{2}) &\approx& m_{\chi}^{2}\,,\qquad\qquad\qquad\qquad\qquad\quad
(p_{1} p_{3}) \approx (p_{2} p_{3}) \approx m_{\chi}^{2}(2-x_{4})\,,\\ 
(p_{1} p_{4}) &\approx& (p_{2} p_{4}) 
\approx m_{\chi}^{2} x_{4}\,,\qquad\qquad\qquad
(p_{1} p_{5}) \approx (p_{2} p_{5}) \approx m_{\chi}^{2} x_{5}\,,\\
(p_{1} p_{6}) &\approx& (p_{2} p_{6}) \approx m_{\chi}^{2}
x_{6}, \qquad\qquad\qquad
p_{3}^{2} \approx 2(p_{5} p_{6}),\\
(p_{4} p_{5}) &\approx& 2 m_{\chi}^{2} (1-x_{6}-\frac{m_{4}^{2}}{4
m_{\chi}^{2}})\,,\qquad \quad\;\;
(p_{4} p_{6}) \approx 2 m_{\chi}^{2}(1-x_{5}+\frac{m_{4}^{2}}{4
m_{\chi}^{2}})\,,\\
(p_{5} p_{6}) &\approx& 2 m_{\chi}^{2}(1-x_{4}+\frac{m_{4}^{2}}{4
m_{\chi}^{2}})\,,
\end{eqnarray*}
and note
\begin{eqnarray}
k &=& p_{4}- p_{2},\\
q &=& 2\frac{(p_{5} p_{1}) p_{6} + (p_{6} p_{1}) p_{5}}
{(p_{5} p_{6})}\,.
\end{eqnarray}

\end{document}